\documentclass[reprint,amsmath,amssymb,aps,pra, superscriptaddress]{revtex4-2}
\usepackage{graphicx}
\usepackage{dcolumn}
\usepackage{bm}
\usepackage[hidelinks]{hyperref}
\hypersetup{colorlinks=true, urlcolor=blue, linkcolor=blue, citecolor=blue}

\begin{document}

%Accoustic Challenges Faced by Antennal Insects
%Antennal-Based Strategies for Sound Localization in Insects
%Sound Localization in Antennal Insects: Challenges and Potential Solutions
\title{Antennal-Based Strategies for Sound Localization by Insects}

\author{Justin Faber}
\email{faber@physics.ucla.edu}
\affiliation{Department of Physics \& Astronomy, University of California, Los Angeles, California, USA}

\author{Alexandros C Alampounti}
\affiliation{University College London Ear Institute: London, UK}

\author{Marcos Georgiades}
\affiliation{University College London Ear Institute: London, UK}

\author{Joerg T Albert}
\affiliation{University College London Ear Institute: London, UK}
\affiliation{Cluster of Excellence Hearing4all, Sensory Physiology \& Behaviour Group, Department for Neuroscience, School of Medicine and Health Sciences, Carl von Ossietzky Universität Oldenburg, Oldenburg, Germany}

\author{Dolores Bozovic}
\affiliation{Department of Physics \& Astronomy, University of California, Los Angeles, California, USA}
\affiliation{California NanoSystems Institute, University of California, Los Angeles, California, USA}

\date{\today}

\begin{abstract}

Insects rely on their hearing in order to communicate, identify and locate potential mates, and avoid predators. Due to their small sizes, many insect species are not able to utilize the interaural time and intensity differences employed by vertebrates for the localization of sound, but have instead evolved other mechanisms to perform this task. One such mechanism is the antenna, which provides directionally sensitive acoustic information. In the current work, we discuss the physical limitations imposed by the Gabor limit and the nature of acoustic radiation as small length scales. We then propose mechanisms that antennal insects may use in order to localize sound and extract precise frequency information from transient signals, thereby circumventing these physical limitations.

\end{abstract}

\maketitle

\begin{figure*}[t]
\includegraphics[width=\textwidth]{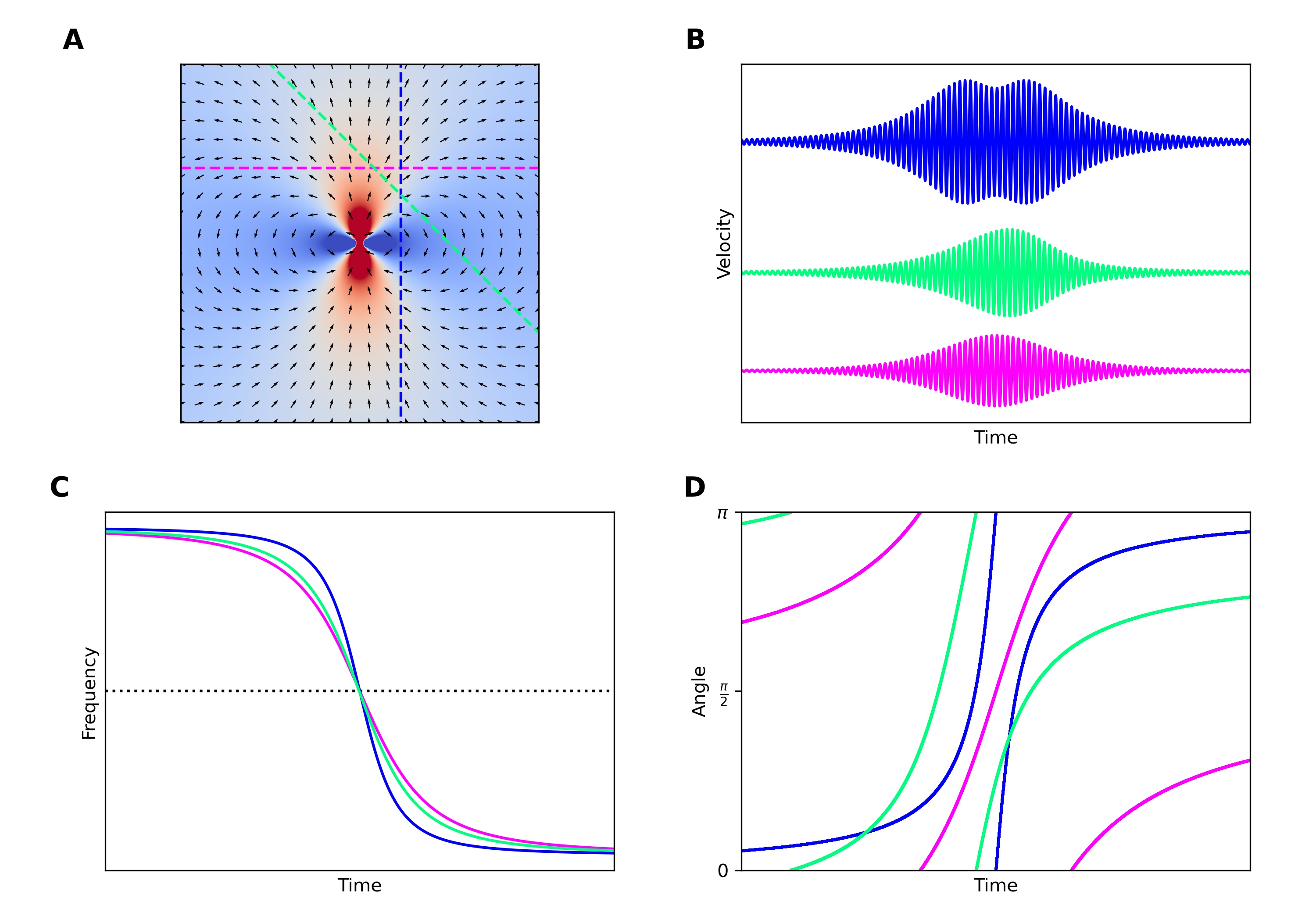}
\caption{(A) Velocity field of an acoustic dipole at one instance in time. The heatmap illustrates dependence of the field strength on the polar angle, with dark red (dark blue) corresponding to the strongest (weakest) regions of the field. The three dashed lines correspond to three potential flight paths of the receiver, traveling from left-to-right and top-to-bottom. (B) Amplitude envelopes of the velocity field, as experienced by the three receivers illustrated in (A). (C) Frequency of the sound experienced by the receivers relative to the frequency emitted by the source (black dotted line). (D) Angle at which the sound is captured by the receiver, relative to receiver's direction of travel, with $\pi$-rotational symmetry.}
\label{fig:vector_field}
\end{figure*}

\textit{Introduction}--Acoustic communication is ubiquitous among vertebrates and insects. Survival and fitness often depend on an organism's ability to detect sound, in order to identify predators, locate mates, and remain in contact with other members of the species. Vertebrates, and many species of insects utilize pressure sensors in the form of tympanic membranes to detect and localize sound \citep{hudspethIntegratingActiveProcess2014, Reichenbach2014}. In vertebrates, directionality is typically inferred from the time and intensity differences of the acoustic waves arriving at the two ears \citep{grotheMechanismsSoundLocalization2010}. These differences, however, are obscured in insects, whose length scale is much shorter than the wavelength of the sound \citep{bennet-clarkAcousticsInsectSong1971, bennet-clarkSizeScaleEffects1998}. In this regime, the acoustic wave is minimally scattered by the body of the receiving insect, leading to approximately the same intensity at the two ears \citep{albertComparativeAspectsHearing2016, romerDirectionalHearingInsects2020}. Further, if the distance between two insect ears is much smaller than the wavelength of sound, the phase difference between the two signals is negligible. Insect species that employ pressure sensors, such as crickets and moths, communicate using high frequencies, thereby avoiding these limitations \citep{bennet-clarkAcousticsInsectSong1971, bennet-clarkSizeScaleEffects1998, barberTigerMothResponses2006}.

Insects that communicate using low frequencies, such as mosquitoes, flies, and bees, employ an entirely different mechanism for localizing sound. They instead use directionally-sensitive antennae \citep{nadrowskiAntennalHearingInsects2011,
windmillMechanicalSpecializationsInsect2016, warrenBridgingGapMammal2021}. The feather-like flagellum pivots in response to variations in the velocity field of the surrounding air particles. Active sensory elements known as \textit{scolopidia} are attached to the base of the flagellum \citep{booFineStructureScolopidia1975, beltonStructureProbableFunction1989}. The scolopidia act as stretch receptors in insects, transmitting neurological signals in response to mechanical deflection of the flagellum \citep{lapshinFrequencyOrganizationJohnstons2017, lapshinDirectionalFrequencyCharacteristics2019}. The direction in which the flagellum pivots corresponds to the direction of the sound source. Each direction is therefore poised to activate a specific subset of the sensory neurons, allowing the insect to identify the direction of the sound source \citep{gopfertNanometrerangeAcousticSensitivity2000, gopfertActiveProcessesInsect2008, gopfertHearingInsects2016}. Flying insects locate sound sources in 3-dimensional space. However, these antennae have, at most, two spatial degrees of freedom. To date, it remains unclear how 3-dimensional spacial information is encoded in the movement of 2-dimensional structures \citep{seoMechanismScalingWing2019}.

In the current work, we first discuss two physical limitations experienced by insects. The first is their short range of detection \citep{feugereMosquitoSoundCommunication2021}, which may arise from the insects' wingbeats acting as acoustic dipole sources \citep{sueurSoundRadiationFlying2005, seoMechanismScalingWing2019}. The second limitation is the Gabor uncertainty principle \citep{gaborAcousticalQuantaTheory1947, cohenTimefrequencyAnalysisTheory1995, oppenheimHumanTimefrequencyAcuity2013}, which imposes upper bounds on the precision at which one can extract frequency and temporal information from a transient signal. We then propose and discuss several potential mechanisms that insects may employ in order to gain additional acoustic information. These mechanisms entail the extraction of information from signal modulations produced during a fly-by event, where two insects enter and then exit each others' detection range. We discuss the expected amplitude, frequency, and phase modulations in the acoustic signal upon arrival at the receiver. Finally, we propose that one or more of these modulations may be used by antennal insects such as mosquitoes, midges, flies, and honeybees in order to reliably detect acoustic signals from conspecifics.

\textit{Range of detection}--It has previously been shown that insects' wingbeats can be approximated by acoustic dipoles \citep{sueurSoundRadiationFlying2005, arthurMosquitoAedesAegypti2014, seoMechanismScalingWing2019}, which produce acoustic fields that falls off rapidly with distance. Further, the feather-like antenna is believed to detect the velocity field produced by a sound source, rather than the pressure field that is sensed by tympanal ears \citep{windmillMechanicalSpecializationsInsect2016}. The combination of the dipole approximation and the detection of particle velocity rather than pressure, suggests that the strength of signals communicated between antennal insects falls off as $1/r^3$, where $r$ is the distance between the sound source and the receiver (see Appendix). By comparison, the monopole-like sources detected by larger animals produce pressure fields that fall off approximately as $1/r$.

This stark difference suggests two things. First, the signal amplitude is very sensitive to changes in distance, due to the inverse cubic dependence. This could potentially be useful for inferring the distance of a sound source, especially when that distance is changing with time. Secondly, one would expect the detection range between antennal insects to be short, as is the case for communication between mosquitoes, which exhibit a detection range of approximately 10 cm \citep{feugereMosquitoSoundCommunication2021}. However, other studies have suggested a much larger range of auditory detection, when sound was emitted by loudspeakers \citep{mendaLongShortHearing2019}. Perhaps the discrepancy can be understood from the different approximations of two types of sound sources. Boxed loudspeakers act as acoustic monopole sources, with velocity fields falling of as $1/r^2$, rather than $1/r^3$.

In FIG. \ref{fig:vector_field} A, we illustrate the velocity field produced by a dipole source at one instance in time. The magnitude of this field is highly sensitive to the distance from the source, and also depends on the orientation of the receiver, relative to the source's direction of travel. The magnitude of the field is hence strongest in front of and behind the insect \citep{sueurSoundRadiationFlying2005, arthurMosquitoAedesAegypti2014, seoMechanismScalingWing2019}, and significantly weaker at other orientations. This results in a wide variety of amplitude envelopes that can be captured during fly-by events (FIG. \ref{fig:vector_field} B). The amplitude envelope can be unimodal, bimodal, and can display an asymmetry, dependent on the angle at which the fly-by event occurs. The shapes of these amplitude envelopes may serve as cues for insects to identify the proximity and orientation of potential mates.

\textit{Gabor limit}--The short range of acoustic communication between insects implies that the signals received from fly-by events are transient. This imposes a limitation on the frequency resolution captured by the receiver. The Gabor uncertainty principle states that an event cannot be localized with arbitrarily high resolution in both the time and frequency domains simultaneously, using Fourier transform pairs \citep{gaborAcousticalQuantaTheory1947, cohenTimefrequencyAnalysisTheory1995}. A transient signal confined to a short period of time requires a wide bandwidth for accurate representation in the frequency domain. Likewise, resolving details confined to a narrow frequency bandwidth requires long periods of time (Fig. \ref{fig:Gabor_limit} of the Appendix).

The Gabor limit states that the representation of any signal is bounded by

\begin{eqnarray}
\Delta t \Delta f \geq \frac{1}{4\pi} \approx 0.08 \text{ cycles},
\label{eqn:delta_f}
\end{eqnarray}

\noindent where $\Delta t$ and $\Delta f$ are the standard deviations of the signal's amplitude envelope in the time and frequency domains, respectively. $\Delta t$ quantifies the signal duration, while $\Delta f$ quantifies its bandwidth. One would therefore expect the frequency resolution used for communication between insects to be limited by the amount of time spent within detection range of each other.

\textit{Doppler shift}--We now consider the possibility of insects using Doppler shifts as cues for locating and identifying the direction of travel of sound sources. In FIG. \ref{fig:vector_field} C, we show the frequency as a function of time, as measured at the receiver, for several fly-by scenarios. The shift in frequency becomes increasingly rapid for closer fly-by events. The instantaneous frequency measured at the receiver can be approximated as

\begin{eqnarray}
f(t) \approx f_0 - \frac{1}{\lambda_0}\frac{dR}{dt},
\label{eqn:phi}
\end{eqnarray}

\noindent provided that the speed of the source, relative to the receiver, is small compared to the speed of sound. $\lambda_0$ and $R$ represent the wavelength of the sound and the receiver's distance from the source, respectively. $f_0$ is the frequency of the sound in the reference frame of the source. We consider the largest Doppler shift possible, which occurs when two insects approach each other head-on and then depart directly away from each other. The difference in frequency between an approaching and departing sound source moving at constant speed is thus $\Delta f = \frac{2}{\lambda_0} |\frac{dR}{dt}|$. The time spent within detection range is $\Delta t = 2R_0/|\frac{dR}{dt}|$, where $R_0$ is the range of detection. Inserting these two quantities into the Gabor limit yields

\begin{eqnarray}
4\pi \Delta t \Delta f = \frac{16\pi R_0}{\lambda_0} \geq 1.
\label{eqn:delta_f}
\end{eqnarray}

\noindent Interestingly, this limitation does not depend on the flight speed of the insects, $|\frac{dR}{dt}|$, as it affects the uncertainty in the time and frequency domains in exactly reciprocal manner. To check if the Gabor limit is satisfied for mosquitoes, we approximate the wavelength of the sound produced by their wingbeats to be $\lambda_0 \approx 1$ meter \citep{suSexSpeciesSpecific2018} and the range of detection to be $R_0 \approx 0.1$ meters \citep{feugereMosquitoSoundCommunication2021}. For these values, estimated from prior experimental studies, we find the Gabor limit to be easily satisfied.

However, an additional, physiological limitation needs to be considered. If one listens to a pure tone for a long period of time, one's precision of the tone's frequency does not increase indefinitely with time, as it would if measured with a digital frequency analyzer. Instead, the limit of frequency precision would rapidly saturate, possibly due to a finite number of receptors. Approximating the flight speed of mosquitoes to be $|\frac{dR}{dt}| \approx 1$ m/s \citep{Chapter1Fastest}, we would expect a frequency shift of $\Delta f \approx 2$ Hz. To detect this frequency shift, a mosquito would need to be able to resolve fractional frequency differences of $\frac{\Delta f}{f_0} \approx \frac{2}{500}$, where we have approximated the flight tone of interest to be $f_0 \approx 500$ Hz \citep{suSexSpeciesSpecific2018}. The level of frequency resolution in insects would hence need to be comparable to that of humans, who can resolve ratios of $\frac{\Delta f}{f_0} \approx \frac{2}{1000}$ \citep{hudspethIntegratingActiveProcess2014}. Note that mosquitoes fly rather slowly compared to other insects. Thus, this shift should be much larger for faster flying insects, such as the dessert locust, which has been measured to fly at approximately ten times the speed of mosquitoes \citep{Chapter1Fastest}. We therefore speculate that the Doppler shift may contain essential information that insects use in order to identify the direction of travel of predators, prey, or potential mates.

\begin{figure*}[t]
\includegraphics[width=\textwidth]{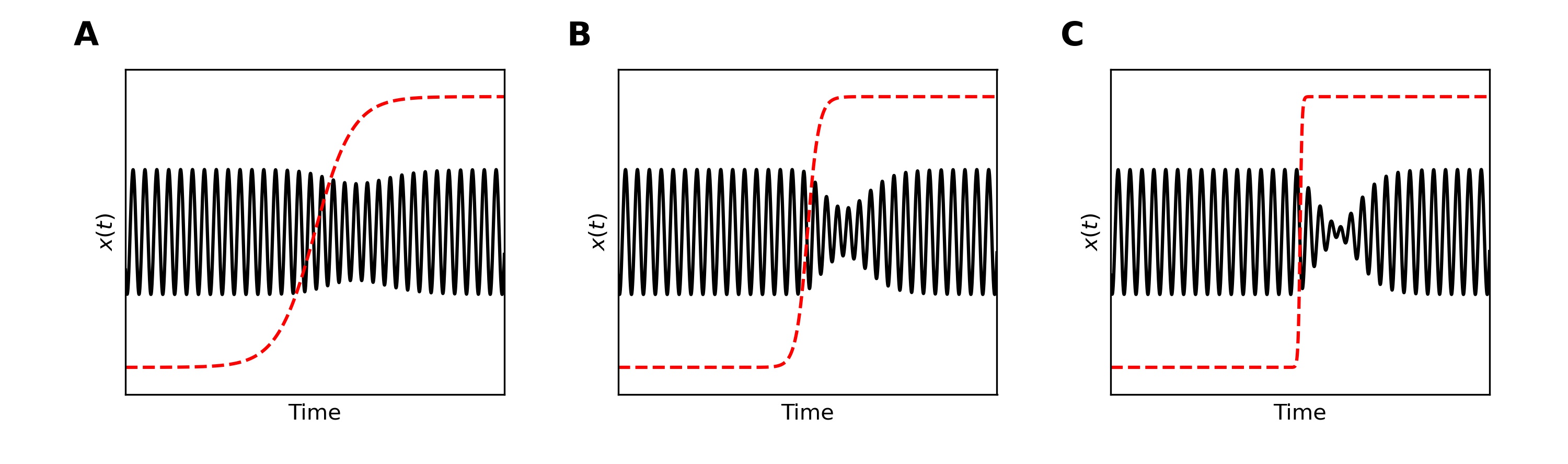}
\caption{Amplitude modulation in the response of a Hopf oscillator to a phase shift in the stimulus waveform. The red, dashed curves show the phase offset as a function of time. The time constant associated with the phase shift was set to $\tau_{ps} = 10, 3, \text{and } 0.5$ for (A-C), respectively.}
\label{fig:phase_shift}
\end{figure*}

\textit{Rapid phase shift}--In FIG. \ref{fig:vector_field} D, we observe that the direction of the sound source, relative to the receiver's direction of travel, changes significantly and rapidly throughout each fly-by event, suggesting that many, if not all, of the sensory elements will be stimulated at some point during the event. We can also note that each fly-by event results in a unique trajectory of the relative angle of the velocity field. Each trajectory would correspond to a unique order and duration in which the sensory elements would be stimulated. The information from the array of sensors could thus be used to infer the position and velocity of the source.

We also consider what information could be obtained by a single sensor. Based on the calculation for the velocity field produced by a flying insect (see Appendix), we should expect the signal in front of the source to be out of phase with the signal behind the source, due to a change in the sign of the $\hat{r}$ component of the velocity field at $\theta=\pm \pi/2$. This half-cycle phase difference has been measured in the flight tones of tethered mosquitoes \citep{arthurMosquitoAedesAegypti2014}. During a fly-by event, we should therefore expect a rapid phase shift in the signal measured by the receiver, in addition to the previously discussed amplitude and frequency modulation. Using a Hopf oscillator as a simple numerical model for auditory detection \citep{choeModelAmplificationHairbundle1998, eguiluzEssentialNonlinearitiesHearing2000} (see Appendix), we demonstrate that phase shifts in the signal of interest yield amplitude modulations in the response of the system, with more rapid phase shifts leading to larger modulations in the response (FIG. \ref{fig:phase_shift}).

\textit{Rapid frequency modulation by mosquitoes}--Mosquitoes rely on precise frequency information for the ability to locate and communicate with conspecifics \citep{wishartFlightResponsesVarious1959, warrenSexRecognitionMidflight2009a, suSexSpeciesSpecific2018}. They have also been observed to display rapid modulations of their wingbeat frequency when in close proximity to a potential mate \citep{simoesRoleAcousticDistortion2016, simoesPrecopulaAcousticBehaviour2017, aldersleyFemaleResistanceHarmonic2019}. It has been proposed that these rapid frequency modulations serve as mating cues between male and female mosquitoes. However, the short acoustic range of mosquito communication imposes a constraint on the duration of the audible signal detectable by mosquitoes in motion. A consequence of the previously discussed Gabor limit is that a short signal duration limits the frequency resolution of detection. We calculate the limitations on frequency resolution of mosquitoes and show that simple, linear responses and transformations to the frequency domain are insufficient for the detection of the rapid frequency modulations exhibited by potential mates. We propose that mosquitoes circumvent this physical limitation by utilizing nonlinear distortion products (DPs) of their response (see Appendix), which are not subject to the Gabor limit. These DPs were previously shown to be essential elements of acoustic communication between mosquitoes \citep{suSexSpeciesSpecific2018, simoesMaskingAuditoryBehaviour2018}.

\begin{figure}[h!]
\includegraphics[width=\columnwidth]{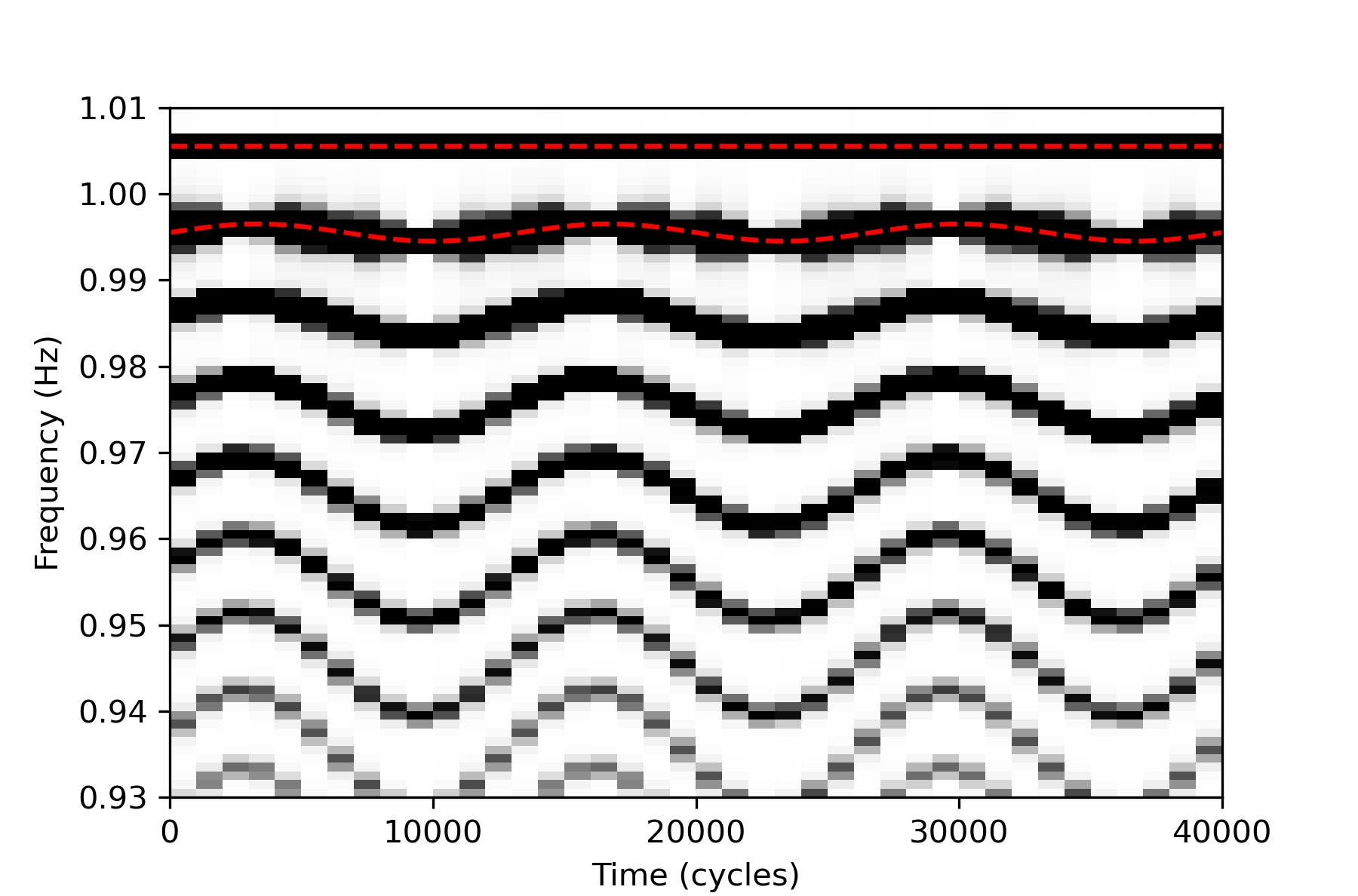}
\caption{Spectrogram of the response of a Hopf oscillator in the quiescent state ($\mu=-0.1$) to a 2-tone stimulus. The red, dashed lines indicate the stimulus tones. Slow modulations of the lower-frequency tone can be seen in the response at this frequency and in the distortion products.}
\label{fig:spec}
\end{figure}

We illustrate the limitation by showing the response of a quiescent Hopf oscillator to a two-tone stimulus, where one of the tones is frequency modulated (FIG. \ref{fig:spec}). The constant-frequency tone represents the detection of the mosquito's own wingbeats. We see that the magnitude of modulation increases with increasing distortion-product order. Every vertical slice of the spectrogram is a Fourier transform of a time window of the signal. We assume non-overlapping time windows; however, these results remain valid for overlapping windows as well. We define the selected window of time to be $T_{window}$,  which contains $N$ samples. The larger the window size, the greater the frequency resolution will be in each vertical slice of the spectrogram. There are $N/2$ positive frequency bins, uniformly spaced from 0 to the Nyquist frequency (half of the sampling rate). Thus, the width of each frequency bin is

\begin{eqnarray}
\Delta f = \frac{f_{sample rate}/2}{N/2} = \frac{1}{T_{window}},
\end{eqnarray}

\noindent which is independent of the sampling rate. Each pixel in the spectrogram has a width of $T_{window}$ and a height of $\Delta f = 1/T_{window}$. When the pixels are compressed in one dimension, either vertically or horizontally, they expand in the other. In fact, pixel area is conserved and unitless: $T_{window} \times \Delta f = 1$.

We consider a frequency-modulated signal of the form

\begin{eqnarray}
\begin{split}
S(t) &= \sin[2\pi f(t)t] \\
f(t) &= f_0 + \frac{\delta f}{2}\sin[2\pi f_{mod}t],
\end{split}
\end{eqnarray}

\noindent where $f(t)$ and $f_0$ represent the instantaneous frequency of the signal and the carrier frequency, respectively. $f_{mod}$ is the frequency of the modulator, and $T_{mod} = 1/f_{mod}$ is the modulation period. To ensure sufficient frequency resolution for capturing the modulations of interest, we need $\delta f \geq 2\Delta f = 2/T_{window}$. However, if the window is too large, the modulations will be averaged over time. We therefore assert at least two pixels per modulation period, the minimum needed to resolve the modulation: $T_{mod} \geq 2T_{window}$. Together, these two inequalities yield the bounds on the time window required to resolve the modulations of interest:

\begin{eqnarray}
\frac{2}{\delta f} \leq T_{window}  \leq \frac{ T_{mod}}{2}.
\end{eqnarray}

When producing a spectrogram, we have the freedom to select a time window that captures the features of interest. However, we consider the limiting case where $\delta f$ is small, or $T_{mod}$ is short. The limit occurs when equality holds among all 3 quantities, and we have only one choice for the time window. For a given modulation period, the smallest fluctuation in frequency that can be resolved is

\begin{eqnarray}
\delta f_{min} = \frac{4}{T_{mod}} = 4f_{mod}.
\label{eq:f_min}
\end{eqnarray}

\noindent If the modulations are weak (small $\delta f$), then the modulation period must be sufficiently long in order to resolve them. Conversely, if the modulations are rapid, then they must have sufficient magnitude to be resolved. This result does not depend on the sampling rate, choice of the time window, or details of the carrier signal.

We now consider the results of a study that showed modulation frequencies to be approximately $f_{mod} \approx 12.5$ Hz  \citep{simoesPrecopulaAcousticBehaviour2017}. According to Eq. \ref{eq:f_min}, the minimum change in frequency that could be resolved at this modulation rate is 
$\delta f_{min} \approx 50$ Hz. However, the study reports peak-to-peak modulation magnitudes ranging from 20 to 200 Hz, many of which fall below our calculated limit. This suggests that if mosquitoes are to detect the rapid frequency modulations of each others' wingbeats, then, even in ideal and noiseless conditions, they may be violating the limitations imposed by the Gabor uncertainty principle.

All of the above calculations consider only the primary tone of the response and linear algorithms for extracting frequency information. Alternatively, mosquitoes may employ a nonlinear algorithm for detecting these rapid frequency modulations, thereby circumventing the Gabor limit. It was previously proposed that human hearing violates this limit and, therefore, must employ nonlinear algorithms \citep{oppenheimHumanTimefrequencyAcuity2013}. 

Mosquitoes have been shown to use nonlinear distortion products to detect flight tones \citep{suSexSpeciesSpecific2018, simoesMaskingAuditoryBehaviour2018}. We therefore propose that detection of nonlinear distortion products allows insects to circumvent the Gabor limit and extract precise frequency information \citep{faberMosquitoInspiredTheoreticalFramework2025}. As shown in Fig. \ref{fig:spec}, these distortion products modulate at larger amplitudes than the primary tones. The frequencies of the distortion products can be expressed as $f_{p, q} = p f_1 - q f_2$, where $f_1$ and $f_2$ the the primary tones, and $p$ and $q$ are integers corresponding to the distortion product order. We can immediately see that the magnitude of the frequency modulations increases with increasing distortion product order. Letting $f_1$ be the modulated tone, we find that the limit given in Eq. \ref{eq:f_min} relaxes with increasing DP order:

\begin{eqnarray}
\delta f_{min} = \frac{4f_{mod}}{p}.
\end{eqnarray}

\noindent By utilizing distortion products generated by the interaction between male and female flight tones, we expect that mosquitoes, and possibly other insects, can resolve changes in frequency that would otherwise be restricted by the Gabor limit.

\textit{Discussion}--Vertebrate animals utilize pressure sensors to detect sound and therefore must have at least two ears, sufficiently separated in space in order to infer the direction of the incoming sound. Directional information is then captured by the interaural time and intensity differences between the two ears \citep{grotheMechanismsSoundLocalization2010}. Additional directional information can be obtained from the asymmetrical structure of the outer ear, which filters sound differently, depending on the direction from which it came.

Insects, however, detect sounds of wavelength much longer than the length scale of their bodies. In this regime, the phase and intensity differences between the two ears is negligible, strictly limiting the utility of pressure sensors. While some insects have evolved to have directionally-sensitive pressure-difference receivers \citep{romerDirectionalHearingInsects2020}, many utilize antennae for detecting and locating sound. Antennal insects therefore face a unique set of challenges not experienced by land vertebrates. They extract information from the vector-valued velocity field of sound, rather than the scalar pressure field. The velocity field naturally contains directional information of the sound source, which is absent in the pressure field. This allows insects to extract the directional information, but with the cost of having a reduced acoustic range of detection, due to the velocity field decaying more rapidly with distance. This small acoustic range in turn results in shorter signals experienced by the sensory systems, and therefore limits frequency resolution.

We have discussed the physical limitations that accompany antennal hearing, namely the short range of detection and the low frequency resolution that arises from the Gabor limit. We then proposed several potential mechanisms that could be employed by antennal insects in order to obtain additional acoustic information. These mechanism entail capturing information contained in the modulations of the acoustic signals. We discussed the transient amplitude-modulated signal experienced by a receiver moving past a sound-emitting insect in flight, and showed that the shape of the amplitude envelope depends heavily on the angle of the source's direction of travel relative to the receiver. We then discussed the Doppler shift and its potential role in sound localization. Additionally, we discussed the rapid phase shifts that arise from the spacial asymmetry in the velocity field of an acoustic dipole. Finally, we discussed the rapid frequency modulations produced by flying mosquitoes and proposed that nonlinear distortion products may be utilized to aid in their detection. We propose that insects may use a combination of these mechanisms in order to overcome the challenges that accompany acoustic detection at small length scales.

\textit{Acknowledgments}--This work was supported by a grant from the Biotechnology and Biological Sciences Research Council, UK (BBSRC, BB/V007866/1 to J.T.A.) and a grant from The Human Frontier Science Program (HFSP grant RGP0033/2021 to J.T.A. and D.B.)

\bibliography{Bibliography}

\begin{thebibliography}{38}%
\makeatletter
\providecommand \@ifxundefined [1]{%
 \@ifx{#1\undefined}
}%
\providecommand \@ifnum [1]{%
 \ifnum #1\expandafter \@firstoftwo
 \else \expandafter \@secondoftwo
 \fi
}%
\providecommand \@ifx [1]{%
 \ifx #1\expandafter \@firstoftwo
 \else \expandafter \@secondoftwo
 \fi
}%
\providecommand \natexlab [1]{#1}%
\providecommand \enquote  [1]{``#1''}%
\providecommand \bibnamefont  [1]{#1}%
\providecommand \bibfnamefont [1]{#1}%
\providecommand \citenamefont [1]{#1}%
\providecommand \href@noop [0]{\@secondoftwo}%
\providecommand \href [0]{\begingroup \@sanitize@url \@href}%
\providecommand \@href[1]{\@@startlink{#1}\@@href}%
\providecommand \@@href[1]{\endgroup#1\@@endlink}%
\providecommand \@sanitize@url [0]{\catcode `\\12\catcode `\$12\catcode
  `\&12\catcode `\#12\catcode `\^12\catcode `\_12\catcode `\%12\relax}%
\providecommand \@@startlink[1]{}%
\providecommand \@@endlink[0]{}%
\providecommand \url  [0]{\begingroup\@sanitize@url \@url }%
\providecommand \@url [1]{\endgroup\@href {#1}{\urlprefix }}%
\providecommand \urlprefix  [0]{URL }%
\providecommand \Eprint [0]{\href }%
\providecommand \doibase [0]{https://doi.org/}%
\providecommand \selectlanguage [0]{\@gobble}%
\providecommand \bibinfo  [0]{\@secondoftwo}%
\providecommand \bibfield  [0]{\@secondoftwo}%
\providecommand \translation [1]{[#1]}%
\providecommand \BibitemOpen [0]{}%
\providecommand \bibitemStop [0]{}%
\providecommand \bibitemNoStop [0]{.\EOS\space}%
\providecommand \EOS [0]{\spacefactor3000\relax}%
\providecommand \BibitemShut  [1]{\csname bibitem#1\endcsname}%
\let\auto@bib@innerbib\@empty
%</preamble>
\bibitem [{\citenamefont
  {Hudspeth}(2014)}]{hudspethIntegratingActiveProcess2014}%
  \BibitemOpen
  \bibfield  {author} {\bibinfo {author} {\bibfnamefont {A.~J.}\ \bibnamefont
  {Hudspeth}},\ }\bibfield  {title} {\bibinfo {title} {Integrating the active
  process of hair cells with cochlear function},\ }\href
  {https://doi.org/10.1038/nrn3786} {\bibfield  {journal} {\bibinfo  {journal}
  {Nat. Rev. Neurosci.}\ }\textbf {\bibinfo {volume} {15}},\ \bibinfo {pages}
  {600} (\bibinfo {year} {2014})}\BibitemShut {NoStop}%
\bibitem [{\citenamefont {Reichenbach}\ and\ \citenamefont
  {Hudspeth}(2014)}]{Reichenbach2014}%
  \BibitemOpen
  \bibfield  {author} {\bibinfo {author} {\bibfnamefont {T.}~\bibnamefont
  {Reichenbach}}\ and\ \bibinfo {author} {\bibfnamefont {a.~J.}\ \bibnamefont
  {Hudspeth}},\ }\bibfield  {title} {\bibinfo {title} {The physics of hearing:
  Fluid mechanics and the active process of the inner ear.},\ }\href
  {https://doi.org/10.1088/0034-4885/77/7/076601} {\bibfield  {journal}
  {\bibinfo  {journal} {Rep. Prog. Phys.}\ }\textbf {\bibinfo {volume} {77}},\
  \bibinfo {pages} {076601} (\bibinfo {year} {2014})}\BibitemShut {NoStop}%
\bibitem [{\citenamefont {Grothe}\ \emph {et~al.}(2010)\citenamefont {Grothe},
  \citenamefont {Pecka},\ and\ \citenamefont
  {McAlpine}}]{grotheMechanismsSoundLocalization2010}%
  \BibitemOpen
  \bibfield  {author} {\bibinfo {author} {\bibfnamefont {B.}~\bibnamefont
  {Grothe}}, \bibinfo {author} {\bibfnamefont {M.}~\bibnamefont {Pecka}},\ and\
  \bibinfo {author} {\bibfnamefont {D.}~\bibnamefont {McAlpine}},\ }\bibfield
  {title} {\bibinfo {title} {Mechanisms of {{Sound Localization}} in
  {{Mammals}}},\ }\href {https://doi.org/10.1152/physrev.00026.2009} {\bibfield
   {journal} {\bibinfo  {journal} {Physiol. Rev.}\ }\textbf {\bibinfo {volume}
  {90}},\ \bibinfo {pages} {983} (\bibinfo {year} {2010})}\BibitemShut
  {NoStop}%
\bibitem [{\citenamefont
  {{Bennet-Clark}}(1971)}]{bennet-clarkAcousticsInsectSong1971}%
  \BibitemOpen
  \bibfield  {author} {\bibinfo {author} {\bibfnamefont {H.~C.}\ \bibnamefont
  {{Bennet-Clark}}},\ }\bibfield  {title} {\bibinfo {title} {Acoustics of
  {{Insect Song}}},\ }\href {https://doi.org/10.1038/234255a0} {\bibfield
  {journal} {\bibinfo  {journal} {Nature}\ }\textbf {\bibinfo {volume} {234}},\
  \bibinfo {pages} {255} (\bibinfo {year} {1971})}\BibitemShut {NoStop}%
\bibitem [{\citenamefont
  {{Bennet-Clark}}(1998)}]{bennet-clarkSizeScaleEffects1998}%
  \BibitemOpen
  \bibfield  {author} {\bibinfo {author} {\bibfnamefont {H.~C.}\ \bibnamefont
  {{Bennet-Clark}}},\ }\bibfield  {title} {\bibinfo {title} {Size and {{Scale
  Effects}} as {{Constraints}} in {{Insect Sound Communication}}},\ }\href@noop
  {} {\bibfield  {journal} {\bibinfo  {journal} {Philos. Trans. Biol. Sci.}\
  }\textbf {\bibinfo {volume} {353}},\ \bibinfo {pages} {407} (\bibinfo {year}
  {1998})}\BibitemShut {NoStop}%
\bibitem [{\citenamefont {Albert}\ and\ \citenamefont
  {Kozlov}(2016)}]{albertComparativeAspectsHearing2016}%
  \BibitemOpen
  \bibfield  {author} {\bibinfo {author} {\bibfnamefont {J.~T.}\ \bibnamefont
  {Albert}}\ and\ \bibinfo {author} {\bibfnamefont {A.~S.}\ \bibnamefont
  {Kozlov}},\ }\bibfield  {title} {\bibinfo {title} {Comparative {{Aspects}} of
  {{Hearing}} in {{Vertebrates}} and {{Insects}} with {{Antennal Ears}}},\
  }\href {https://doi.org/10.1016/j.cub.2016.09.017} {\bibfield  {journal}
  {\bibinfo  {journal} {Curr. Biol.}\ }\textbf {\bibinfo {volume} {26}},\
  \bibinfo {pages} {R1050} (\bibinfo {year} {2016})}\BibitemShut {NoStop}%
\bibitem [{\citenamefont
  {R{\"o}mer}(2020)}]{romerDirectionalHearingInsects2020}%
  \BibitemOpen
  \bibfield  {author} {\bibinfo {author} {\bibfnamefont {H.}~\bibnamefont
  {R{\"o}mer}},\ }\bibfield  {title} {\bibinfo {title} {Directional hearing in
  insects: Biophysical, physiological and ecological challenges},\ }\href
  {https://doi.org/10.1242/jeb.203224} {\bibfield  {journal} {\bibinfo
  {journal} {J. Exp. Biol.}\ }\textbf {\bibinfo {volume} {223}},\ \bibinfo
  {pages} {jeb203224} (\bibinfo {year} {2020})}\BibitemShut {NoStop}%
\bibitem [{\citenamefont {Barber}\ and\ \citenamefont
  {Conner}(2006)}]{barberTigerMothResponses2006}%
  \BibitemOpen
  \bibfield  {author} {\bibinfo {author} {\bibfnamefont {J.~R.}\ \bibnamefont
  {Barber}}\ and\ \bibinfo {author} {\bibfnamefont {W.~E.}\ \bibnamefont
  {Conner}},\ }\bibfield  {title} {\bibinfo {title} {Tiger moth responses to a
  simulated bat attack: Timing and duty cycle},\ }\href
  {https://doi.org/10.1242/jeb.02295} {\bibfield  {journal} {\bibinfo
  {journal} {J. Exp. Biol.}\ }\textbf {\bibinfo {volume} {209}},\ \bibinfo
  {pages} {2637} (\bibinfo {year} {2006})}\BibitemShut {NoStop}%
\bibitem [{\citenamefont {Nadrowski}\ \emph {et~al.}(2011)\citenamefont
  {Nadrowski}, \citenamefont {Effertz}, \citenamefont {Senthilan},\ and\
  \citenamefont {G{\"o}pfert}}]{nadrowskiAntennalHearingInsects2011}%
  \BibitemOpen
  \bibfield  {author} {\bibinfo {author} {\bibfnamefont {B.}~\bibnamefont
  {Nadrowski}}, \bibinfo {author} {\bibfnamefont {T.}~\bibnamefont {Effertz}},
  \bibinfo {author} {\bibfnamefont {P.~R.}\ \bibnamefont {Senthilan}},\ and\
  \bibinfo {author} {\bibfnamefont {M.~C.}\ \bibnamefont {G{\"o}pfert}},\
  }\bibfield  {title} {\bibinfo {title} {Antennal hearing in insects - {{New}}
  findings, new questions},\ }\href
  {https://doi.org/10.1016/j.heares.2010.03.092} {\bibfield  {journal}
  {\bibinfo  {journal} {Hear. Res.}\ }\textbf {\bibinfo {volume} {273}},\
  \bibinfo {pages} {7} (\bibinfo {year} {2011})}\BibitemShut {NoStop}%
\bibitem [{\citenamefont {Windmill}\ and\ \citenamefont
  {Jackson}(2016)}]{windmillMechanicalSpecializationsInsect2016}%
  \BibitemOpen
  \bibfield  {author} {\bibinfo {author} {\bibfnamefont {J.~F.~C.}\
  \bibnamefont {Windmill}}\ and\ \bibinfo {author} {\bibfnamefont {J.~C.}\
  \bibnamefont {Jackson}},\ }\bibfield  {title} {\bibinfo {title} {Mechanical
  {{Specializations}} of {{Insect Ears}}},\ }in\ \href
  {https://doi.org/10.1007/978-3-319-28890-1_6} {\emph {\bibinfo {booktitle}
  {Insect {{Hearing}}}}},\ \bibinfo {editor} {edited by\ \bibinfo {editor}
  {\bibfnamefont {G.~S.}\ \bibnamefont {Pollack}}, \bibinfo {editor}
  {\bibfnamefont {A.~C.}\ \bibnamefont {Mason}}, \bibinfo {editor}
  {\bibfnamefont {A.~N.}\ \bibnamefont {Popper}},\ and\ \bibinfo {editor}
  {\bibfnamefont {R.~R.}\ \bibnamefont {Fay}}}\ (\bibinfo  {publisher}
  {Springer International Publishing},\ \bibinfo {year} {2016})\ pp.\ \bibinfo
  {pages} {125--157}\BibitemShut {NoStop}%
\bibitem [{\citenamefont {Warren}\ and\ \citenamefont
  {Nowotny}(2021)}]{warrenBridgingGapMammal2021}%
  \BibitemOpen
  \bibfield  {author} {\bibinfo {author} {\bibfnamefont {B.}~\bibnamefont
  {Warren}}\ and\ \bibinfo {author} {\bibfnamefont {M.}~\bibnamefont
  {Nowotny}},\ }\bibfield  {title} {\bibinfo {title} {Bridging the {{Gap
  Between Mammal}} and {{Insect Ears}} -- {{A Comparative}} and {{Evolutionary
  View}} of {{Sound-Reception}}},\ }\href
  {https://doi.org/10.3389/fevo.2021.667218} {\bibfield  {journal} {\bibinfo
  {journal} {Front. Ecol. Evol.}\ }\textbf {\bibinfo {volume} {9}},\ \bibinfo
  {pages} {1} (\bibinfo {year} {2021})}\BibitemShut {NoStop}%
\bibitem [{\citenamefont {Boo}\ and\ \citenamefont
  {Richards}(1975)}]{booFineStructureScolopidia1975}%
  \BibitemOpen
  \bibfield  {author} {\bibinfo {author} {\bibfnamefont {K.~S.}\ \bibnamefont
  {Boo}}\ and\ \bibinfo {author} {\bibfnamefont {A.~G.}\ \bibnamefont
  {Richards}},\ }\bibfield  {title} {\bibinfo {title} {Fine structure of the
  scolopidia in the johnston's organ of male {{Aedes}} aegypti ({{L}}.)
  ({{Diptera}}: {{Culicidae}})},\ }\href
  {https://doi.org/10.1016/0020-7322(75)90031-8} {\bibfield  {journal}
  {\bibinfo  {journal} {Int. J. of Insect Morphol. Embryol.}\ }\textbf
  {\bibinfo {volume} {4}},\ \bibinfo {pages} {549} (\bibinfo {year}
  {1975})}\BibitemShut {NoStop}%
\bibitem [{\citenamefont {Belton}(1989)}]{beltonStructureProbableFunction1989}%
  \BibitemOpen
  \bibfield  {author} {\bibinfo {author} {\bibfnamefont {P.}~\bibnamefont
  {Belton}},\ }\bibfield  {title} {\bibinfo {title} {The structure and probable
  function of the internal cuticular parts of {{Johnston}}'s organ in
  mosquitoes ({{Aedes}} aegypti)},\ }\href {https://doi.org/10.1139/z89-371}
  {\bibfield  {journal} {\bibinfo  {journal} {Can. J. Zool.}\ }\textbf
  {\bibinfo {volume} {67}},\ \bibinfo {pages} {2625} (\bibinfo {year}
  {1989})}\BibitemShut {NoStop}%
\bibitem [{\citenamefont {Lapshin}\ and\ \citenamefont
  {Vorontsov}(2017)}]{lapshinFrequencyOrganizationJohnstons2017}%
  \BibitemOpen
  \bibfield  {author} {\bibinfo {author} {\bibfnamefont {D.~N.}\ \bibnamefont
  {Lapshin}}\ and\ \bibinfo {author} {\bibfnamefont {D.~D.}\ \bibnamefont
  {Vorontsov}},\ }\bibfield  {title} {\bibinfo {title} {Frequency organization
  of the {{Johnston}}'s organ in male mosquitoes ({{Diptera}},
  {{Culicidae}})},\ }\href {https://doi.org/10.1242/jeb.152017} {\bibfield
  {journal} {\bibinfo  {journal} {J. Exp.l Biol.}\ }\textbf {\bibinfo {volume}
  {220}},\ \bibinfo {pages} {3927} (\bibinfo {year} {2017})}\BibitemShut
  {NoStop}%
\bibitem [{\citenamefont {Lapshin}\ and\ \citenamefont
  {Vorontsov}(2019)}]{lapshinDirectionalFrequencyCharacteristics2019}%
  \BibitemOpen
  \bibfield  {author} {\bibinfo {author} {\bibfnamefont {D.~N.}\ \bibnamefont
  {Lapshin}}\ and\ \bibinfo {author} {\bibfnamefont {D.~D.}\ \bibnamefont
  {Vorontsov}},\ }\bibfield  {title} {\bibinfo {title} {Directional and
  frequency characteristics of auditory neurons in {{Culex}} male mosquitoes},\
  }\href {https://doi.org/10.1242/jeb.208785} {\bibfield  {journal} {\bibinfo
  {journal} {J. Exp. Biol.}\ }\textbf {\bibinfo {volume} {222}},\ \bibinfo
  {pages} {1} (\bibinfo {year} {2019})}\BibitemShut {NoStop}%
\bibitem [{\citenamefont {G{\"o}pfert}\ and\ \citenamefont
  {Robert}(2000)}]{gopfertNanometrerangeAcousticSensitivity2000}%
  \BibitemOpen
  \bibfield  {author} {\bibinfo {author} {\bibfnamefont {M.~C.}\ \bibnamefont
  {G{\"o}pfert}}\ and\ \bibinfo {author} {\bibfnamefont {D.}~\bibnamefont
  {Robert}},\ }\bibfield  {title} {\bibinfo {title} {Nanometre-range acoustic
  sensitivity in male and female mosquitoes},\ }\href
  {https://doi.org/10.1098/rspb.2000.1021} {\bibfield  {journal} {\bibinfo
  {journal} {Proc. Roy. Soc. London, Ser. B, Biol. Sci.}\ }\textbf {\bibinfo
  {volume} {267}},\ \bibinfo {pages} {453} (\bibinfo {year}
  {2000})}\BibitemShut {NoStop}%
\bibitem [{\citenamefont {G{\"o}pfert}\ and\ \citenamefont
  {Robert}(2008)}]{gopfertActiveProcessesInsect2008}%
  \BibitemOpen
  \bibfield  {author} {\bibinfo {author} {\bibfnamefont {M.~C.}\ \bibnamefont
  {G{\"o}pfert}}\ and\ \bibinfo {author} {\bibfnamefont {D.}~\bibnamefont
  {Robert}},\ }\bibfield  {title} {\bibinfo {title} {Active {{Processes}} in
  {{Insect Hearing}}},\ }in\ \href
  {https://doi.org/10.1007/978-0-387-71469-1_6} {\emph {\bibinfo {booktitle}
  {Active {{Processes}} and {{Otoacoustic Emissions}} in {{Hearing}}}}},\
  \bibinfo {series and number} {Springer {{Handbook}} of {{Auditory
  Research}}},\ \bibinfo {editor} {edited by\ \bibinfo {editor} {\bibfnamefont
  {G.~A.}\ \bibnamefont {Manley}}, \bibinfo {editor} {\bibfnamefont {R.~R.}\
  \bibnamefont {Fay}},\ and\ \bibinfo {editor} {\bibfnamefont {A.~N.}\
  \bibnamefont {Popper}}}\ (\bibinfo  {publisher} {Springer},\ \bibinfo
  {address} {New York, NY},\ \bibinfo {year} {2008})\ pp.\ \bibinfo {pages}
  {191--209}\BibitemShut {NoStop}%
\bibitem [{\citenamefont {G{\"o}pfert}\ and\ \citenamefont
  {Hennig}(2016)}]{gopfertHearingInsects2016}%
  \BibitemOpen
  \bibfield  {author} {\bibinfo {author} {\bibfnamefont {M.~C.}\ \bibnamefont
  {G{\"o}pfert}}\ and\ \bibinfo {author} {\bibfnamefont {R.~M.}\ \bibnamefont
  {Hennig}},\ }\bibfield  {title} {\bibinfo {title} {Hearing in {{Insects}}},\
  }\href {https://doi.org/10.1146/annurev-ento-010715-023631} {\bibfield
  {journal} {\bibinfo  {journal} {Annu. Rev. Entomol.}\ }\textbf {\bibinfo
  {volume} {61}},\ \bibinfo {pages} {257} (\bibinfo {year} {2016})}\BibitemShut
  {NoStop}%
\bibitem [{\citenamefont {Seo}\ \emph {et~al.}(2019)\citenamefont {Seo},
  \citenamefont {Hedrick},\ and\ \citenamefont
  {Mittal}}]{seoMechanismScalingWing2019}%
  \BibitemOpen
  \bibfield  {author} {\bibinfo {author} {\bibfnamefont {J.-H.}\ \bibnamefont
  {Seo}}, \bibinfo {author} {\bibfnamefont {T.~L.}\ \bibnamefont {Hedrick}},\
  and\ \bibinfo {author} {\bibfnamefont {R.}~\bibnamefont {Mittal}},\
  }\bibfield  {title} {\bibinfo {title} {Mechanism and scaling of wing tone
  generation in mosquitoes},\ }\href {https://doi.org/10.1088/1748-3190/ab54fc}
  {\bibfield  {journal} {\bibinfo  {journal} {Bioinspir. Biomim.}\ }\textbf
  {\bibinfo {volume} {15}},\ \bibinfo {pages} {016008} (\bibinfo {year}
  {2019})}\BibitemShut {NoStop}%
\bibitem [{\citenamefont {Feug{\`e}re}\ \emph {et~al.}(2021)\citenamefont
  {Feug{\`e}re}, \citenamefont {Gibson}, \citenamefont {Manoukis},\ and\
  \citenamefont {Roux}}]{feugereMosquitoSoundCommunication2021}%
  \BibitemOpen
  \bibfield  {author} {\bibinfo {author} {\bibfnamefont {L.}~\bibnamefont
  {Feug{\`e}re}}, \bibinfo {author} {\bibfnamefont {G.}~\bibnamefont {Gibson}},
  \bibinfo {author} {\bibfnamefont {N.~C.}\ \bibnamefont {Manoukis}},\ and\
  \bibinfo {author} {\bibfnamefont {O.}~\bibnamefont {Roux}},\ }\bibfield
  {title} {\bibinfo {title} {Mosquito sound communication: Are male swarms loud
  enough to attract females?},\ }\href {https://doi.org/10.1098/rsif.2021.0121}
  {\bibfield  {journal} {\bibinfo  {journal} {J. R. Soc. Interface}\ }\textbf
  {\bibinfo {volume} {18}},\ \bibinfo {pages} {20210121} (\bibinfo {year}
  {2021})}\BibitemShut {NoStop}%
\bibitem [{\citenamefont {Sueur}\ \emph {et~al.}(2005)\citenamefont {Sueur},
  \citenamefont {Tuck},\ and\ \citenamefont
  {Robert}}]{sueurSoundRadiationFlying2005}%
  \BibitemOpen
  \bibfield  {author} {\bibinfo {author} {\bibfnamefont {J.}~\bibnamefont
  {Sueur}}, \bibinfo {author} {\bibfnamefont {E.~J.}\ \bibnamefont {Tuck}},\
  and\ \bibinfo {author} {\bibfnamefont {D.}~\bibnamefont {Robert}},\
  }\bibfield  {title} {\bibinfo {title} {Sound radiation around a flying fly},\
  }\href {https://doi.org/10.1121/1.1932227} {\bibfield  {journal} {\bibinfo
  {journal} {J. Acoust. Soc. Am.}\ }\textbf {\bibinfo {volume} {118}},\
  \bibinfo {pages} {530} (\bibinfo {year} {2005})}\BibitemShut {NoStop}%
\bibitem [{\citenamefont {Gabor}(1947)}]{gaborAcousticalQuantaTheory1947}%
  \BibitemOpen
  \bibfield  {author} {\bibinfo {author} {\bibfnamefont {D.}~\bibnamefont
  {Gabor}},\ }\bibfield  {title} {\bibinfo {title} {Acoustical {{Quanta}} and
  the {{Theory}} of {{Hearing}}},\ }\href {https://doi.org/10.1038/159591a0}
  {\bibfield  {journal} {\bibinfo  {journal} {Nature}\ }\textbf {\bibinfo
  {volume} {159}},\ \bibinfo {pages} {591} (\bibinfo {year}
  {1947})}\BibitemShut {NoStop}%
\bibitem [{\citenamefont {Cohen}(1995)}]{cohenTimefrequencyAnalysisTheory1995}%
  \BibitemOpen
  \bibfield  {author} {\bibinfo {author} {\bibfnamefont {L.}~\bibnamefont
  {Cohen}},\ }\href@noop {} {\emph {\bibinfo {title} {Time-Frequency
  Analysis}}}\ (\bibinfo  {publisher} {Prentice-Hall PTR},\ \bibinfo {year}
  {1995})\BibitemShut {NoStop}%
\bibitem [{\citenamefont {Oppenheim}\ and\ \citenamefont
  {Magnasco}(2013)}]{oppenheimHumanTimefrequencyAcuity2013}%
  \BibitemOpen
  \bibfield  {author} {\bibinfo {author} {\bibfnamefont {J.~N.}\ \bibnamefont
  {Oppenheim}}\ and\ \bibinfo {author} {\bibfnamefont {M.~O.}\ \bibnamefont
  {Magnasco}},\ }\bibfield  {title} {\bibinfo {title} {Human time-frequency
  acuity beats the fourier uncertainty principle},\ }\bibfield  {journal}
  {\bibinfo  {journal} {Phys. Rev. Lett.}\ }\textbf {\bibinfo {volume} {110}},\
  \href {https://doi.org/10.1103/PhysRevLett.110.044301}
  {10.1103/PhysRevLett.110.044301} (\bibinfo {year} {2013}),\ \Eprint
  {https://arxiv.org/abs/1208.4611} {1208.4611} \BibitemShut {NoStop}%
\bibitem [{\citenamefont {Arthur}\ \emph {et~al.}(2014)\citenamefont {Arthur},
  \citenamefont {Emr}, \citenamefont {Wyttenbach},\ and\ \citenamefont
  {Hoy}}]{arthurMosquitoAedesAegypti2014}%
  \BibitemOpen
  \bibfield  {author} {\bibinfo {author} {\bibfnamefont {B.~J.}\ \bibnamefont
  {Arthur}}, \bibinfo {author} {\bibfnamefont {K.~S.}\ \bibnamefont {Emr}},
  \bibinfo {author} {\bibfnamefont {R.~A.}\ \bibnamefont {Wyttenbach}},\ and\
  \bibinfo {author} {\bibfnamefont {R.~R.}\ \bibnamefont {Hoy}},\ }\bibfield
  {title} {\bibinfo {title} {Mosquito ( {{Aedes}} aegypti ) flight tones:
  {{Frequency}}, harmonicity, spherical spreading, and phase relationships},\
  }\href {https://doi.org/10.1121/1.4861233} {\bibfield  {journal} {\bibinfo
  {journal} {J. Acoust. Soc. Am.}\ }\textbf {\bibinfo {volume} {135}},\
  \bibinfo {pages} {933} (\bibinfo {year} {2014})}\BibitemShut {NoStop}%
\bibitem [{\citenamefont {Menda}\ \emph {et~al.}(2019)\citenamefont {Menda},
  \citenamefont {Nitzany}, \citenamefont {Shamble}, \citenamefont {Wells},
  \citenamefont {Harrington}, \citenamefont {Miles},\ and\ \citenamefont
  {Hoy}}]{mendaLongShortHearing2019}%
  \BibitemOpen
  \bibfield  {author} {\bibinfo {author} {\bibfnamefont {G.}~\bibnamefont
  {Menda}}, \bibinfo {author} {\bibfnamefont {E.~I.}\ \bibnamefont {Nitzany}},
  \bibinfo {author} {\bibfnamefont {P.~S.}\ \bibnamefont {Shamble}}, \bibinfo
  {author} {\bibfnamefont {A.}~\bibnamefont {Wells}}, \bibinfo {author}
  {\bibfnamefont {L.~C.}\ \bibnamefont {Harrington}}, \bibinfo {author}
  {\bibfnamefont {R.~N.}\ \bibnamefont {Miles}},\ and\ \bibinfo {author}
  {\bibfnamefont {R.~R.}\ \bibnamefont {Hoy}},\ }\bibfield  {title} {\bibinfo
  {title} {The {{Long}} and {{Short}} of {{Hearing}} in the {{Mosquito Aedes}}
  aegypti},\ }\href {https://doi.org/10.1016/j.cub.2019.01.026} {\bibfield
  {journal} {\bibinfo  {journal} {Curr. Biol.}\ }\textbf {\bibinfo {volume}
  {29}},\ \bibinfo {pages} {709} (\bibinfo {year} {2019})}\BibitemShut
  {NoStop}%
\bibitem [{\citenamefont {Su}\ \emph {et~al.}(2018)\citenamefont {Su},
  \citenamefont {Andr{\'e}s}, \citenamefont {{Boyd-Gibbins}}, \citenamefont
  {Somers},\ and\ \citenamefont {Albert}}]{suSexSpeciesSpecific2018}%
  \BibitemOpen
  \bibfield  {author} {\bibinfo {author} {\bibfnamefont {M.~P.}\ \bibnamefont
  {Su}}, \bibinfo {author} {\bibfnamefont {M.}~\bibnamefont {Andr{\'e}s}},
  \bibinfo {author} {\bibfnamefont {N.}~\bibnamefont {{Boyd-Gibbins}}},
  \bibinfo {author} {\bibfnamefont {J.}~\bibnamefont {Somers}},\ and\ \bibinfo
  {author} {\bibfnamefont {J.~T.}\ \bibnamefont {Albert}},\ }\bibfield  {title}
  {\bibinfo {title} {Sex and species specific hearing mechanisms in mosquito
  flagellar ears},\ }\bibfield  {journal} {\bibinfo  {journal} {Nat. Commun.}\
  }\textbf {\bibinfo {volume} {9}},\ \href
  {https://doi.org/10.1038/s41467-018-06388-7} {10.1038/s41467-018-06388-7}
  (\bibinfo {year} {2018})\BibitemShut {NoStop}%
\bibitem [{Cha()}]{Chapter1Fastest}%
  \BibitemOpen
  \href@noop {} {\bibinfo {title} {Chapter 1: {{Fastest Flyer}} {\textbar}
  {{The University}} of {{Florida Book}} of {{Insect Records}} {\textbar}
  {{Department}} of {{Entomology}} \& {{Nematology}} {\textbar}
  {{UF}}/{{IFAS}}}}\BibitemShut {NoStop}%
\bibitem [{\citenamefont {Choe}\ \emph {et~al.}(1998)\citenamefont {Choe},
  \citenamefont {Magnasco},\ and\ \citenamefont
  {Hudspeth}}]{choeModelAmplificationHairbundle1998}%
  \BibitemOpen
  \bibfield  {author} {\bibinfo {author} {\bibfnamefont {Y.}~\bibnamefont
  {Choe}}, \bibinfo {author} {\bibfnamefont {M.~O.}\ \bibnamefont {Magnasco}},\
  and\ \bibinfo {author} {\bibfnamefont {A.~J.}\ \bibnamefont {Hudspeth}},\
  }\bibfield  {title} {\bibinfo {title} {A model for amplification of
  hair-bundle motion by cyclical binding of {{Ca2}}+ to
  mechanoelectrical-transduction channels},\ }\href
  {https://doi.org/10.1073/pnas.95.26.15321} {\bibfield  {journal} {\bibinfo
  {journal} {PNAS}\ }\textbf {\bibinfo {volume} {95}},\ \bibinfo {pages}
  {15321} (\bibinfo {year} {1998})}\BibitemShut {NoStop}%
\bibitem [{\citenamefont {Egu{\'i}luz}\ \emph {et~al.}(2000)\citenamefont
  {Egu{\'i}luz}, \citenamefont {Ospeck}, \citenamefont {Choe}, \citenamefont
  {Hudspeth},\ and\ \citenamefont
  {Magnasco}}]{eguiluzEssentialNonlinearitiesHearing2000}%
  \BibitemOpen
  \bibfield  {author} {\bibinfo {author} {\bibfnamefont {V.~M.}\ \bibnamefont
  {Egu{\'i}luz}}, \bibinfo {author} {\bibfnamefont {M.}~\bibnamefont {Ospeck}},
  \bibinfo {author} {\bibfnamefont {Y.}~\bibnamefont {Choe}}, \bibinfo {author}
  {\bibfnamefont {A.~J.}\ \bibnamefont {Hudspeth}},\ and\ \bibinfo {author}
  {\bibfnamefont {M.~O.}\ \bibnamefont {Magnasco}},\ }\bibfield  {title}
  {\bibinfo {title} {Essential {{Nonlinearities}} in {{Hearing}}},\ }\href
  {https://doi.org/10.1103/PhysRevLett.84.5232} {\bibfield  {journal} {\bibinfo
   {journal} {Phys. Rev. Lett.}\ }\textbf {\bibinfo {volume} {84}},\ \bibinfo
  {pages} {5232} (\bibinfo {year} {2000})}\BibitemShut {NoStop}%
\bibitem [{\citenamefont {Wishart}\ and\ \citenamefont
  {Riordan}(1959)}]{wishartFlightResponsesVarious1959}%
  \BibitemOpen
  \bibfield  {author} {\bibinfo {author} {\bibfnamefont {G.}~\bibnamefont
  {Wishart}}\ and\ \bibinfo {author} {\bibfnamefont {D.~F.}\ \bibnamefont
  {Riordan}},\ }\bibfield  {title} {\bibinfo {title} {Flight {{Responses}} to
  {{Various Sounds}} by {{Adult Males}} of aedes aegypti (l.) (diptera :
  Culicidae)},\ }\href {https://doi.org/10.4039/Ent91181-3} {\bibfield
  {journal} {\bibinfo  {journal} {Can. Entomol.}\ }\textbf {\bibinfo {volume}
  {91}},\ \bibinfo {pages} {181} (\bibinfo {year} {1959})}\BibitemShut
  {NoStop}%
\bibitem [{\citenamefont {Warren}\ \emph {et~al.}(2009)\citenamefont {Warren},
  \citenamefont {Gibson},\ and\ \citenamefont
  {Russell}}]{warrenSexRecognitionMidflight2009a}%
  \BibitemOpen
  \bibfield  {author} {\bibinfo {author} {\bibfnamefont {B.}~\bibnamefont
  {Warren}}, \bibinfo {author} {\bibfnamefont {G.}~\bibnamefont {Gibson}},\
  and\ \bibinfo {author} {\bibfnamefont {I.~J.}\ \bibnamefont {Russell}},\
  }\bibfield  {title} {\bibinfo {title} {Sex {{Recognition}} through
  {{Midflight Mating Duets}} in {{Culex Mosquitoes Is Mediated}} by {{Acoustic
  Distortion}}},\ }\href {https://doi.org/10.1016/j.cub.2009.01.059} {\bibfield
   {journal} {\bibinfo  {journal} {Curr. Biol.}\ }\textbf {\bibinfo {volume}
  {19}},\ \bibinfo {pages} {485} (\bibinfo {year} {2009})}\BibitemShut
  {NoStop}%
\bibitem [{\citenamefont {Sim{\~o}es}\ \emph {et~al.}(2016)\citenamefont
  {Sim{\~o}es}, \citenamefont {Ingham}, \citenamefont {Gibson},\ and\
  \citenamefont {Russell}}]{simoesRoleAcousticDistortion2016}%
  \BibitemOpen
  \bibfield  {author} {\bibinfo {author} {\bibfnamefont {P.~M.~V.}\
  \bibnamefont {Sim{\~o}es}}, \bibinfo {author} {\bibfnamefont {R.~A.}\
  \bibnamefont {Ingham}}, \bibinfo {author} {\bibfnamefont {G.}~\bibnamefont
  {Gibson}},\ and\ \bibinfo {author} {\bibfnamefont {I.~J.}\ \bibnamefont
  {Russell}},\ }\bibfield  {title} {\bibinfo {title} {A role for acoustic
  distortion in novel rapid frequency modulation behaviour in free-flying male
  mosquitoes},\ }\href {https://doi.org/10.1242/jeb.135293} {\bibfield
  {journal} {\bibinfo  {journal} {J. Exp. Biol.}\ }\textbf {\bibinfo {volume}
  {219}},\ \bibinfo {pages} {2039} (\bibinfo {year} {2016})}\BibitemShut
  {NoStop}%
\bibitem [{\citenamefont {Sim{\~o}es}\ \emph {et~al.}(2017)\citenamefont
  {Sim{\~o}es}, \citenamefont {Gibson},\ and\ \citenamefont
  {Russell}}]{simoesPrecopulaAcousticBehaviour2017}%
  \BibitemOpen
  \bibfield  {author} {\bibinfo {author} {\bibfnamefont {P.~M.~V.}\
  \bibnamefont {Sim{\~o}es}}, \bibinfo {author} {\bibfnamefont
  {G.}~\bibnamefont {Gibson}},\ and\ \bibinfo {author} {\bibfnamefont {I.~J.}\
  \bibnamefont {Russell}},\ }\bibfield  {title} {\bibinfo {title} {Pre-copula
  acoustic behaviour of males in the malarial mosquitoes {{Anopheles}} coluzzii
  and {{Anopheles}} gambiae s.s. does not contribute to reproductive
  isolation},\ }\href {https://doi.org/10.1242/jeb.149757} {\bibfield
  {journal} {\bibinfo  {journal} {J. Exp. Biol.}\ }\textbf {\bibinfo {volume}
  {220}},\ \bibinfo {pages} {379} (\bibinfo {year} {2017})}\BibitemShut
  {NoStop}%
\bibitem [{\citenamefont {Aldersley}\ and\ \citenamefont
  {Cator}(2019)}]{aldersleyFemaleResistanceHarmonic2019}%
  \BibitemOpen
  \bibfield  {author} {\bibinfo {author} {\bibfnamefont {A.}~\bibnamefont
  {Aldersley}}\ and\ \bibinfo {author} {\bibfnamefont {L.~J.}\ \bibnamefont
  {Cator}},\ }\bibfield  {title} {\bibinfo {title} {Female resistance and
  harmonic convergence influence male mating success in {{Aedes}} aegypti},\
  }\href {https://doi.org/10.1038/s41598-019-38599-3} {\bibfield  {journal}
  {\bibinfo  {journal} {Sci. Rep.}\ }\textbf {\bibinfo {volume} {9}},\ \bibinfo
  {pages} {2145} (\bibinfo {year} {2019})}\BibitemShut {NoStop}%
\bibitem [{\citenamefont {Sim{\~o}es}\ \emph {et~al.}(2018)\citenamefont
  {Sim{\~o}es}, \citenamefont {Ingham}, \citenamefont {Gibson},\ and\
  \citenamefont {Russell}}]{simoesMaskingAuditoryBehaviour2018}%
  \BibitemOpen
  \bibfield  {author} {\bibinfo {author} {\bibfnamefont {P.~M.}\ \bibnamefont
  {Sim{\~o}es}}, \bibinfo {author} {\bibfnamefont {R.}~\bibnamefont {Ingham}},
  \bibinfo {author} {\bibfnamefont {G.}~\bibnamefont {Gibson}},\ and\ \bibinfo
  {author} {\bibfnamefont {I.~J.}\ \bibnamefont {Russell}},\ }\bibfield
  {title} {\bibinfo {title} {Masking of an auditory behaviour reveals how male
  mosquitoes use distortion to detect females},\ }\href
  {https://doi.org/10.1098/rspb.2017.1862} {\bibfield  {journal} {\bibinfo
  {journal} {Proc. Roy. Soc. B, Biol. Sci.}\ }\textbf {\bibinfo {volume}
  {285}},\ \bibinfo {pages} {11} (\bibinfo {year} {2018})}\BibitemShut
  {NoStop}%
\bibitem [{\citenamefont {Faber}\ \emph {et~al.}(2025)\citenamefont {Faber},
  \citenamefont {Alampounti}, \citenamefont {Georgiades}, \citenamefont
  {Albert},\ and\ \citenamefont
  {Bozovic}}]{faberMosquitoInspiredTheoreticalFramework2025}%
  \BibitemOpen
  \bibfield  {author} {\bibinfo {author} {\bibfnamefont {J.}~\bibnamefont
  {Faber}}, \bibinfo {author} {\bibfnamefont {A.~C.}\ \bibnamefont
  {Alampounti}}, \bibinfo {author} {\bibfnamefont {M.}~\bibnamefont
  {Georgiades}}, \bibinfo {author} {\bibfnamefont {J.~T.}\ \bibnamefont
  {Albert}},\ and\ \bibinfo {author} {\bibfnamefont {D.}~\bibnamefont
  {Bozovic}},\ }\bibfield  {title} {\bibinfo {title} {A {{Mosquito-Inspired
  Theoretical Framework}} for {{Acoustic Signal Detection}}},\ }\bibfield
  {journal} {\bibinfo  {journal} {arXiv}\ }\href
  {https://doi.org/10.48550/arXiv.2501.05576} {10.48550/arXiv.2501.05576}
  (\bibinfo {year} {2025})\BibitemShut {NoStop}%
\bibitem [{\citenamefont {Howe}(2014)}]{howeAcousticsAerodynamicSound2014}%
  \BibitemOpen
  \bibfield  {author} {\bibinfo {author} {\bibfnamefont {M.}~\bibnamefont
  {Howe}},\ }\href {https://doi.org/10.1017/CBO9781107360273} {\emph {\bibinfo
  {title} {Acoustics and {{Aerodynamic Sound}}}}}\ (\bibinfo  {publisher}
  {Cambridge University Press},\ \bibinfo {year} {2014})\BibitemShut {NoStop}%
\end{thebibliography}%

\subsection*{End Matter}

\textit{Appendix A: Velocity field of an acoustic dipole}--To determine the acoustic field produced by a monopole at the origin, we must solve the wave equation,

\begin{eqnarray}
\frac{1}{c^2}\frac{\partial^2 \Phi}{\partial^2 t} - \nabla^2 \Phi = -q(t)\delta(\vec{\bf{x}}),
\label{eqn:wave}
\end{eqnarray}

\noindent where $q(t)$ represents an arbitrary function related to the motion that generates the sound, $c$ is the speed of sound in the material, and $\Phi(\vec{\bf{x}}, t)$ is the velocity potential \citep{howeAcousticsAerodynamicSound2014}. The velocity potential is particularly useful to work with because both the pressure and velocity fields can be calculated using

\begin{eqnarray}
p = -\rho \frac{\partial \Phi}{\partial t} \quad \text{and} \quad \vec{\bf{u}} = -\nabla \Phi,
\label{eqn:p_u}
\end{eqnarray}

\noindent respectively. $\rho$ represents the density of the material. Eq. \ref{eqn:wave} has solution,

\begin{eqnarray}
\Phi_{mon}(r, t) = \frac{q(t - \frac{r}{c})}{4\pi r},
\label{eqn:mono_sol}
\end{eqnarray}

\noindent where $r$ is the distance from the origin. Notice that the solution is spherically symmetric and is delayed by time, $\frac{r}{c}$, which accounts for the finite speed of sound in this compressible material.

To find the velocity potential of a dipole, we must solve

\begin{eqnarray}
\frac{1}{c^2}\frac{\partial^2 \Phi}{\partial^2 t} - \nabla^2 \Phi = -\frac{\partial}{\partial z} [q(t)\delta(\vec{\bf{x}})],
\label{eqn:wave2}
\end{eqnarray}

\noindent which has solution,

\begin{eqnarray}
\Phi_{dip}(r, \theta, t) = \frac{\cos\theta q(t - \frac{r}{c})}{4\pi r^2} +\frac{\cos\theta \frac{dq}{dt}(t - \frac{r}{c})}{4\pi c r},
\label{eqn:dip}
\end{eqnarray}

\noindent where the head and direction of travel of the insect is oriented at $\theta=0$ \citep{howeAcousticsAerodynamicSound2014}. We consider sinusoidal sound waves with 

\begin{eqnarray}
q(t - \frac{r}{c}) = Q e^{i\omega (t - \frac{r}{c})}
\label{eqn:Q}
\end{eqnarray}

\noindent Inserting this into Eq. \ref{eqn:dip}, we find

\begin{eqnarray}
\Phi_{dip}(r, \theta, t) = \frac{\cos\theta Q e^{i\omega(t - \frac{r}{c})}}{4\pi r^2} [1 + \frac{ir}{2\pi\lambda}]
\label{eqn:}
\end{eqnarray}

\noindent We can then calculate the velocity field using $\vec{\bf{u}} = -\nabla \Phi$. We find

\begin{eqnarray}
\begin{split}
\vec{\bf{u}}_{dip} (r, \theta, t) = \frac{Q e^{i\omega(t - \frac{r}{c})}}{4\pi r^3} \bigg[   \bigg( 2 &+ \frac{ir}{\pi\lambda} - \Big(\frac{r}{2\pi\lambda}\Big)^2   \bigg) \cos\theta\hat{r}   \\ &+  \bigg(  1 + \frac{ir}{2\pi\lambda} \bigg)\sin\theta\hat{\theta}  \bigg]
\end{split}
\label{eqn:udip}
\end{eqnarray}

\noindent In the limit where $\frac{r}{2\pi\lambda} \ll 1$ and $\frac{a}{r} \ll 1$ (where $a$ is the dipole separation length scale, or the distance between the two wings of the sound-emitting insect), we find

\begin{eqnarray}
\vec{\bf{u}}_{dip}(r, \theta, t) = \frac{Q e^{i\omega(t - \frac{r}{c})}}{4\pi r^3} [2\cos\theta\hat{r} + \sin\theta\hat{\theta}   ].
\label{eqn:udip}
\end{eqnarray}

\begin{figure}[h!]
\includegraphics[width=\columnwidth]{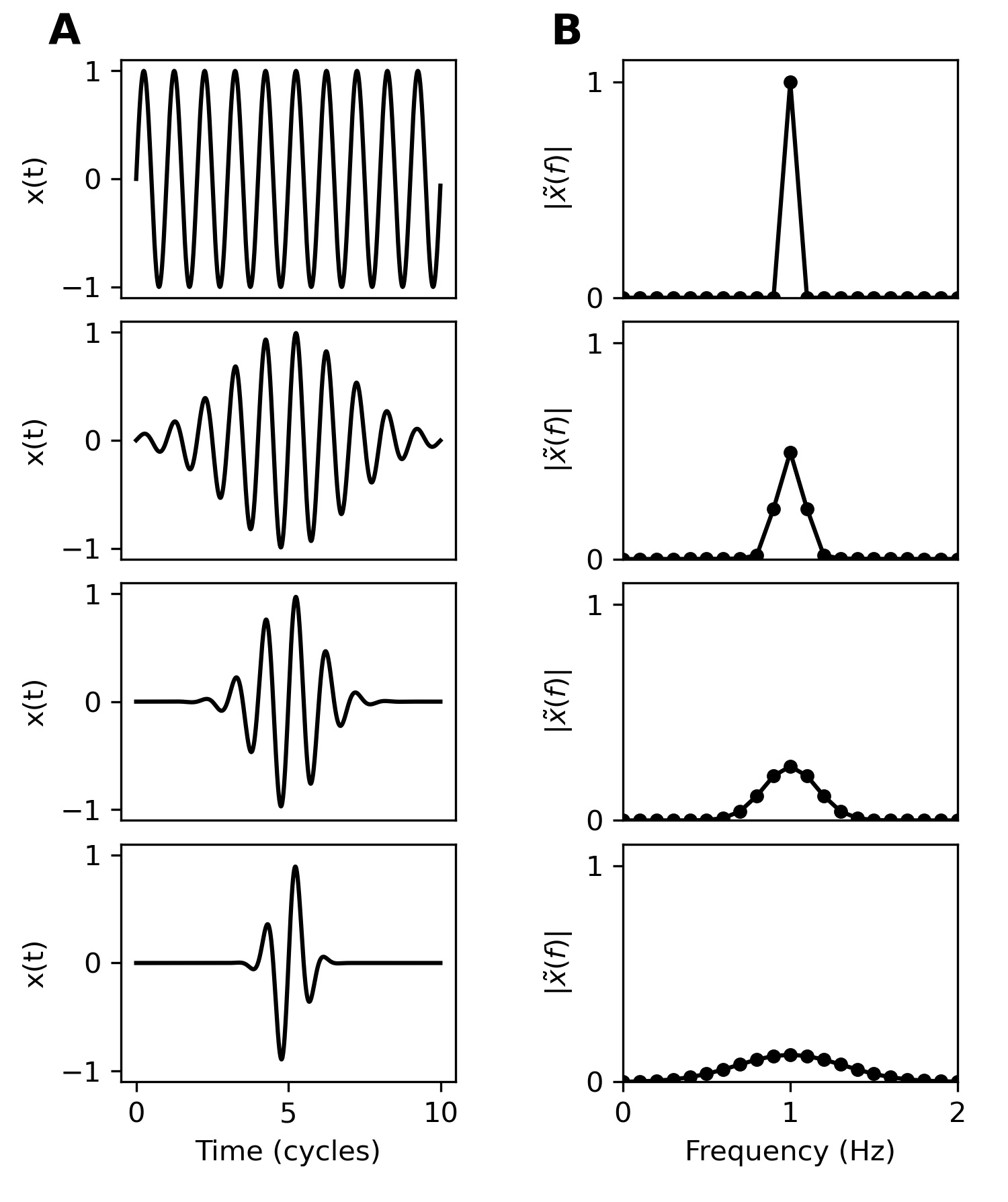}
\caption{(A) Sinusoidal signals with Gaussian amplitude envelopes in the time domain. The signals are each 10 periods in duration and were sampled at 100 samples/period. (B) Magnitude of the Fourier transforms of each of the signals in (A). Data points correspond to individual frequency bins.}
\label{fig:Gabor_limit}
\end{figure}

\textit{Appendix B: Gabor limit}--Here we illustrate the Gabor limit by showing several signals in both the time and frequency domains (Fig. \ref{fig:Gabor_limit}). A signal confined to a short window of time requires a broadband representation in the frequency domain.

\textit{Appendix C: Hopf oscillator}--The Hopf oscillator consists of time-dependent complex variable, $z(t)$. The dynamics of this variable are governed by the normal form equation for Hopf bifurcations \citep{choeModelAmplificationHairbundle1998, eguiluzEssentialNonlinearitiesHearing2000},

\begin{eqnarray}
\frac{dz}{dt} = (\mu + i\omega_0)z - |z|^2z + F(t),
\end{eqnarray}

\noindent where $\omega_0$ and $\mu$ represent the characteristic frequency and control parameter of the detector, respectively. The system resides in the quiescent state for $\mu<0$. The system produces active, autonomous oscillations when $\mu>0$. The boundary between these two regimes ($\mu=0$) is the Hopf bifurcation.

To measure the response of the system to a rapid phase shift, we choose a forcing term of the form,

\begin{eqnarray}
F(t) = f_0 e^{i[\omega t + \phi(t)]},
\end{eqnarray}

\noindent where $f_0$ and $\omega$ represent the stimulus amplitude and frequency, respectively. The time-dependent phase offset takes the form

\begin{eqnarray}
\phi(t) = \frac{\pi}{1 + e^{-\frac{t}{\tau_{ps}}}},
\end{eqnarray}

\noindent where $\tau_{ps}$ is the time constant associated with the speed of the phase shift. We set $\omega = \omega_0 = 1$, $\mu = f_0 = 0.01$, and vary $\tau_{ps}$ to see the effects of phase shifts on the response.

Distortion products are produced in the response when two or more tones are simultaneously presented. To demonstrate this nonlinear response, we choose a forcing term of the form,

\begin{eqnarray}
F(t) = f_1 e^{i\omega_1 t} + f_2 e^{i\omega_2 t},
\end{eqnarray}

\noindent where the forcing amplitudes are represented by $f_1$ and $f_2$, and the forcing frequencies are $\omega_1$ and $\omega_2$. The response of the system is largest at the two stimulus tones. However, nonlinear responses can also be found at the distortion-product frequencies, $f_{p, q} = p f_1 - q f_2$, where $p$ and $q$ are integers corresponding to the distortion-product order. The magnitude of these distortion products falls off with increasing order.

\clearpage

\end{document}